\documentclass[11pt]{article}
\usepackage{DGfest,epsfig}
\usepackage{times}
\input colordvi 

\def\be{\begin{equation}}
\def\ee{\end{equation}}
\def\bea{\begin{eqnarray}}
\def\eea{\end{eqnarray}}

\begin{document}
\baselineskip 11.5pt
\title{FUN WITH DIRAC EIGENVALUES}

\author{ MICHAEL CREUTZ }

\address{Department of Physics, Brookhaven National Laboratory\\
Upton, NY 11973, USA}

%

\maketitle\abstracts{ It is popular to discuss low energy physics in
lattice gauge theory in terms of the small eigenvalues of the lattice
Dirac operator.  I play with some ensuing pitfalls in the
interpretation of these eigenvalue spectra.}

\section{Introduction}

Amongst the lattice gauge community it has recently become quite
popular to study the distributions of eigenvalues of the Dirac
operator in the presence of the background gauge fields generated in
simulations.  There are a variety of motivations for this.  First, in
a classic work, Banks and Casher\cite{Banks:1979yr} related the
density of small Dirac eigenvalues to spontaneous chiral symmetry
breaking.  Second, lattice discretizations of the Dirac operator based
the Ginsparg-Wilson relation\cite{Ginsparg:1981bj} have the
corresponding eigenvalues on circles in the complex plane.  The
validity of various approximations to such an operator can be
qualitatively assessed by looking at the eigenvalues.  Third, using
the overlap method\cite{Neuberger:1997fp} to construct a Dirac
operator with good chiral symmetry has difficulties if the starting
Wilson fermion operator has small eigenvalues.  This can influence the
selection of simulation parameters, such as the gauge
action.\cite{Aoki:2002vt} Finally, since low eigenvalues impede
conjugate gradient methods, separating out these eigenvalues
explicitly can potentially be useful in developing dynamical
simulation algorithms.\cite{Duncan:1998gq}

Despite this interest in the eigenvalue distributions, there are some
dangers inherent in interpreting the observations.  Physical results
come from the full path integral over both the bosonic and fermionic
fields.  Doing these integrals one at a time is fine, but trying to
interpret the intermediate results is inherently dangerous.  While the
Dirac eigenvalues depend on the given gauge field, it is important to
remember that in a dynamical simulation the gauge field distribution
itself depends on the eigenvalues.  This circular behavior gives a
highly non-linear system, and such systems are notoriously hard to
interpret.

Given that this is a joyous occasion, I will present some of this
issues in terms of an amusing set of puzzles arising from naive
interpretations of Dirac eigenvalues on the lattice.  The discussion
is meant to be a mixture of thought provoking and confusing.  It is
not necessarily particularly deep or new.

\section{The framework}

To get started, I need to establish the context of the discussion.  I
consider a generic path integral for a gauge theory
\begin{equation}
Z=\int (dA)(d\psi)(d\overline\psi)\ e^{-S_G(A)+\overline\psi D(A) \psi}.
\end{equation}
Here $A$ and $\psi$ represent the gauge and quark fields,
respectively, $S_G(A)$ is the pure gauge part of the action, and
$D(A)$ represents the Dirac operator in use for the quarks.  As the
action is quadratic in the fermion fields, a formal integration gives
\begin{equation}
Z=\int (dA)\ |D(A)|\ e^{-S_G(A)}.
\label{path}
\end{equation}
Working on a finite, lattice $D(A)$ is a finite dimensional matrix,
and for a given gauge field I can formally consider its eigenvectors
and eigenvalues
\begin{equation}
D(A)\psi_i=\lambda_i\psi_i.
\end{equation}
The determinant appearing in Eq.~(\ref{path}) is the product of these
eigenvalues; so, the path integral takes the form
\begin{equation}
Z=\int (dA)\ e^{-S_G(A)}\ \prod_i \lambda_i.
\end{equation}
Averaging over gauge fields defines the eigenvalue density
\begin{equation}
\rho(x+iy)={1\over {N}Z}\int (dA)\ |D(A)|\ e^{-S_G(A)}
\sum_{i}\delta(x-{\rm Re}\lambda_i(A))\delta(y-{\rm Im}\lambda_i(A)).
\end{equation}
Here $N$ is the dimension of the Dirac operator, including volume,
gauge, spin, and flavor indices.  

In situations where the fermion determinant is not positive, $\rho$
can be negative or complex.  Nevertheless, I still refer to it as a
density.  I will assume that $\rho$ is real; situations where this is
not true, such as with a finite chemical
potential,\cite{Osborn:2005ss} are beyond the scope of this
discussion.

\begin{figure*}
\centering
\includegraphics[width=3in]{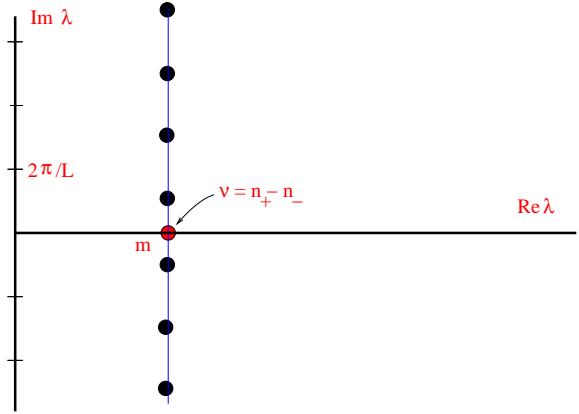}
\caption{ In the naive continuum picture, all eigenvalues of the Dirac
operator lie along a line parallel to the imaginary axis.  In a finite
volume these eigenvalues become discrete.  The real eigenvalues divide
into distinct chiralities and define a topological invariant.
\label{continuum}
}
\end{figure*} 

At zero chemical potential, all actions used in practice satisfy
``$\gamma_5$ hermiticity''
\begin{equation}
\gamma_5 D \gamma_5=D^\dagger.
\label{hermiticity}
\end{equation}
With this condition all non-real eigenvalues occur in complex
conjugate pairs, implying for the density
\begin{equation}
\rho(z)=\rho(z^*).
\end{equation}
This property will be shared by all the operators considered in the
following discussion.  

The quest is to find general statements relating the behavior of the
eigenvalue density to physical properties of the theory.  I repeat the
earlier warning; $\rho$ depends on the distribution of gauge fields
$A$ which in turn is weighted by $\rho$ which depends on the
distribution of $A$ \ldots.

\subsection{The continuum}

Of course the continuum theory is only really defined as the limit of
the lattice theory.  Nevertheless, it is perhaps useful to recall the
standard picture, where the Dirac operator
$$
D=\gamma_\mu (\partial_\mu+igA_\mu)+m
$$
is the sum of an anti-hermitian piece and the quark mass $m$.  All
eigenvalues have the same real part $m$
$$
\rho(x+iy)=\delta(x-m) \tilde\rho(y).
$$ The eigenvalues lie along a line parallel to the imaginary axis,
while the hermiticity condition of Eq.~(\ref{hermiticity}) implies
they occur in complex conjugate pairs.

Restricted to the subspace of real eigenvalues, $\gamma_5$ commutes
with $D$ and thus these eigenvectors can be separated by chirality.
The difference between the number of positive and negative eigenvalues
of $\gamma_5$ in this subspace defines an index related to the
topological structure of the gauge fields.\cite{index} The basic
structure is sketched in Fig.~(\ref{continuum}).

The Banks and Casher argument relates a non-vanishing $\tilde\rho(0)$
to the chiral condensate occurring when the mass goes to zero.  I will
say more on this later in the lattice context.

Note that the naive picture suggests a symmetry between positive and
negative mass.  Due to anomalies, this is spurious.  With an odd
number of flavors, the theory obtained by flipping the signs of all
fermion masses is physically inequivalent to the initial theory.

\begin{figure*}
\centering
\includegraphics[width=2.5in]{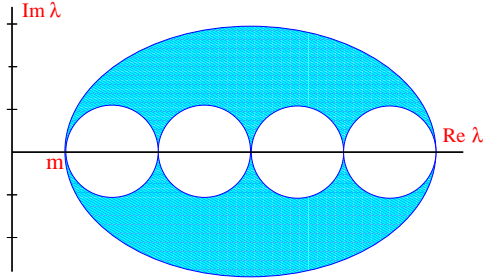}
\caption{ Free Wilson fermions display an eigenvalue spectrum with a
momentum dependent real part.  This removes doublers by giving them a
large effective mass.}
\label{fig2}
\end{figure*} 

\subsection{Wilson fermions}

The lattice reveals that the true situation is considerably more
intricate due to the chiral anomaly.  Without ultraviolet infinities,
all naive symmetries of the lattice action are true symmetries.  Naive
fermions cannot have anomalies, which are cancelled by extra states
referred to as doublers.  Wilson fermions\cite{Wilson:1975id} avoid
the this issue by giving a large real part to those eigenvalues
corresponding to the doublers.  For free Wilson fermions the
eigenvalue structure displays a simple pattern as shown in
Fig.~(\ref{fig2}).

As the gauge fields are turned on, this pattern will fuzz out.  An
additional complication is that the operator $D$ is no longer normal,
i.e. $[D,D^\dagger]\ne 0$ and the eigenvectors need not be orthogonal.
The complex eigenvalues are still paired, although, as the gauge
fields vary, complex pairs of eigenvalues can collide and separate
along the real axis.  In general, the real eigenvalues will form a
continuous distribution.

As in the continuum, an index can be defined from the spectrum of the
Wilson-Dirac operator.  Again, $\gamma_5$ hermiticity allows real
eigenvalues to be sorted by chirality.  To remove the contribution of
the doubler eigenvalues, select a point inside the leftmost open
circle of Fig.~(\ref{fig2}).  Then define the index of the gauge field
to be the net chirality of all real eigenvalues below that point.  For
smooth gauge fields this agrees with the topological winding number
obtained from their interpolation to the continuum.  It also
corresponds to the winding number discussed below for the overlap
operator.

\subsection{The overlap}

\begin{figure*}
\centering
\includegraphics[width=5in]{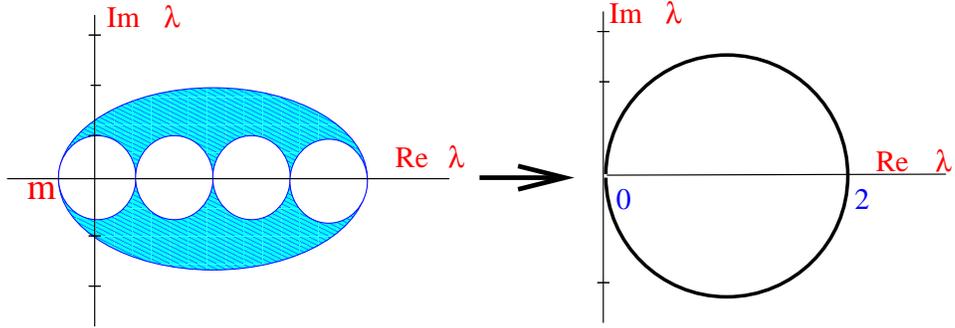}
\caption{The overlap operator is constructed by projecting the Wilson
  Dirac operator onto a unitary operator.}
\label{fig3}
\end{figure*}

Wilson fermions have a rather complicated behavior under chiral
transformations.  The overlap formalism\cite{Neuberger:1997fp}
simplifies this by first projecting the Wilson matrix $D_W$ onto a
unitary operator
\begin{equation}
V=(D_W D_W^\dagger)^{-1/2} D_W.
\end{equation}
This is to be understood in terms of going to a basis that
diagonalizes $D_W D_W^\dagger$, doing the inversion, and then
returning to the initial basis.  In terms of this unitary quantity,
the overlap matrix is
\begin{equation}
D=1+V.
\end{equation}
The projection process is sketched in Fig.~(\ref{fig3}).  The mass
used in the starting Wilson operator is taken to a negative value so
selected that the low momentum states are projected to low
eigenvalues, while the doubler states are driven towards $\lambda\sim
2$.

The overlap operator has several nice properties.  First, it satisfies
the Ginsparg-Wilson relation,\cite{Ginsparg:1981bj} most succinctly
written as the unitarity of $V$ coupled with its $\gamma_5$
hermiticity
\begin{equation}
\gamma_5 V\gamma_5 V=1.
\end{equation}
As it is constructed from a unitary operator, normality of $D$ is
guaranteed.  But, most important, it exhibits a lattice version of an
exact chiral symmetry.\cite{Luscher:1998pq} The fermionic action
$\overline\psi D\psi$ is invariant under the transformation
\begin{eqnarray}
&\psi\rightarrow e^{i\theta\gamma_5}\psi\cr
&\overline\psi\rightarrow \overline\psi e^{i\theta\hat\gamma_5}
\label{symmetry}
\end{eqnarray}
where
\begin{equation}
\hat\gamma_5 =V\gamma_5.
\end{equation}
As with $\gamma_5$, this quantity is Hermitean and its square is unity.
Thus its eigenvalues are all plus or minus unity.  The trace
defines an index
\begin{equation}
\nu={1\over 2}{\rm Tr}\hat\gamma_5
\end{equation}
which plays exactly the role of the index in the continuum.

It is important to note that the overlap operator is not unique.  Its
precise form depends on the particular initial operator chosen to
project onto the unitary form.  Using the Wilson-Dirac operator for
this purpose, the result still depends on the input mass used.  From
its historical origins in the domain wall formalism, this quantity is
sometimes called the ``domain wall height.''

Because the overlap is not unique, an ambiguity can remain in
determining the winding number of a given gauge configuration.  Issues
arise when $D_W D_W^\dagger$ is not invertible, and for a given gauge
field this can occur at specific values of the projection point.
This problem can be avoided for ``smooth'' gauge fields.  Indeed, an
``admissibility condition,''~\cite{Luscher:1981zq,Hernandez:1998et}
requiring all plaquette values to remain sufficiently close to the
identity, removes the ambiguity.  Unfortunately this condition is
incompatible with reflection positivity.\cite{Creutz:2004ir} Because
of these issues, it is not known if the topological susceptibility is
in fact a well defined physical observable.  On the other hand, as it
is not clear how to measure the susceptibility in a scattering
experiment, there seems to be little reason to care if it is an
observable or not.

\begin{figure*}
\centering
\includegraphics[width=3in]{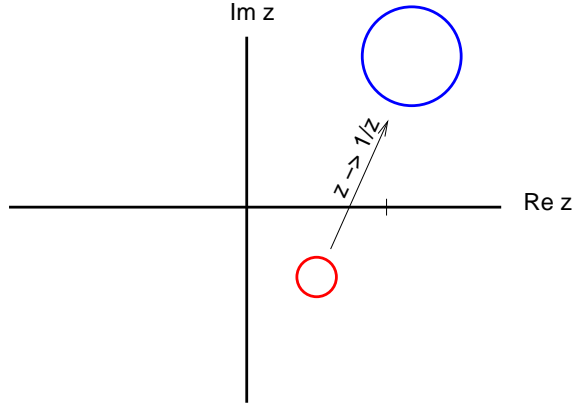}
\caption{Inverting a complex circle generates another circle.}
\label{circles}
\end{figure*} 

\section{A Cheshire chiral condensate}

Now that I have reviewed the basic framework, it is time for a little
fun.  I will calculate the chiral condensate in the overlap formalism.
I should warn you that, in the interest of amusing you, I start the
argument in an intentionally deceptive manner.

\subsection{He's here}

I begin with the standard massless overlap theory.  I want to
calculate the quantity $\langle\overline\psi\psi\rangle$.  Remarkably,
this can be done exactly.  I start with
\begin{equation}
\langle\overline\psi\psi\rangle=\langle {\rm Tr} D^{-1}\rangle
=\left\langle \sum_i {1\over \lambda_i}\right\rangle
=\left\langle \sum {\rm Re} {1\over \lambda_i}\right\rangle
\end{equation}
where I have used the complex pairing of eigenvalues to cancel the
imaginary parts.  At the end, the average is to be taken over
appropriately weighted gauge configurations.

Now the crucial feature of the overlap operator is that its
eigenvalues all lie on a circle in the complex plane.  An interesting
property of a general complex circle is that the inverses of all its
points generates another circle, as sketched in Fig.~\ref{circles}.

\begin{figure*}
\centering
\includegraphics[width=3in]{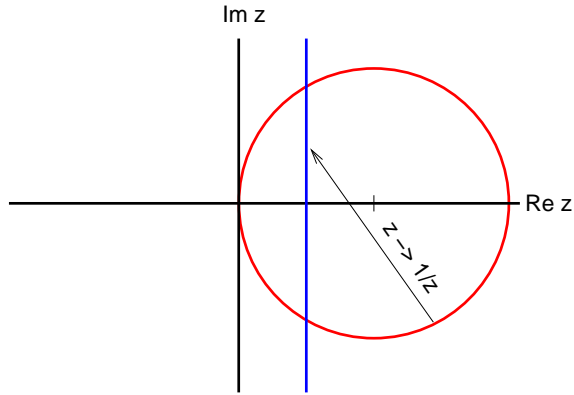}
\caption{Inverting the overlap operator generates a line with real
  part 1/2.}
\label{circles1}
\end{figure*}

This process is, however, somewhat singular for the overlap operator
itself since the corresponding circle touches the origin.  In this
case the inverted circle has infinite radius, i.e. it degenerates into
a line.  For the circle of the overlap operator, with center at $z=1$
and radius 1, the inverse circle is a line with real part 1/2 and
parallel to the imaginary axis.  This is sketched in
Fig.~\ref{circles1}.

This placement of eigenvalues enables an immediate calculation of the
condensate 
\begin{equation}
\langle\overline\psi\psi\rangle=
\sum {\rm Re} {1\over \lambda_i}=\sum {1\over 2}= {N\over 2}.
\end{equation}
Here $N$ is the dimension of the matrix, and includes the expected
volume factor.

So the condensate, supposedly a signal for spontaneous chiral symmetry
breaking, does not vanish!  But something is fishy, I didn't use any
dynamics.  The result also is independent of gauge configuration.

\subsection{He's gone}

So lets get more sophisticated.  On the lattice, the chiral symmetry
is more complicated than in the continuum, involving both $\gamma_5$
and $\hat\gamma_5$ in a rather intricate way.  In particular, the
operator $\overline\psi\psi$ does not transform in any simple manner
under chiral rotations.  A possibly nicer combination is
$\overline\psi(1-D/2)\psi$.  If I consider the rotation in
Eq.~(\ref{symmetry}) with $\theta=\pi/2$, this quantity becomes its
negative.  But it is also easy to calculate the expectation of this as
well.  The second term involves
\begin{equation}
\langle\overline\psi D \psi\rangle={\rm Tr} D^{-1} D={\rm Tr}I=N.
\end{equation}
Putting the two pieces together
\begin{equation}
\langle\overline\psi(1-D/2)\psi\rangle=N/2-N/2=0.
\end{equation}
So, I've lost the chiral condensate that I so easily showed didn't
vanish just a moment ago.  Where did it go?

\begin{figure*}
\centering
\includegraphics[width=3in]{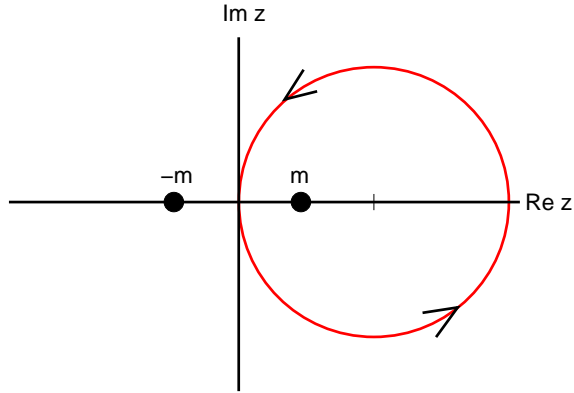}
\caption{As the mass changes sign a pole moves between inside and outside
the overlap circle.  This generates a jump in the condensate.}
\label{circles2}
\end{figure*} 

\subsection{He's back}

The issue lies in a careless treatment of limits.  In finite volume,
$\langle\overline\psi(1-D/2)\psi\rangle$ must vanish just from the
exact lattice chiral symmetry.  This vanishing occurs for all gauge
configurations.  To proceed, introduce a small mass and take the
volume to infinity first and then the mass to zero.  Toward this end,
consider the quantity
\begin{equation}
\langle\overline\psi\psi\rangle=\sum_i {1\over\lambda_i+m}.
\end{equation}
The signal for chiral symmetry breaking is a jump in this quantity as
the mass passes through zero.

As the volume goes to infinity, replace the above sum with a contour
integral around the overlap circle using $z=1+e^{i\theta}$.  Up to the
trivial volume factor, I should evaluate
\begin{equation}
i\int_0^{2\pi} d\theta {\rho(\theta)\over 1+e^{i\theta}+m}.
\end{equation}
As the mass passes through zero, the pole at $z=-m$ passes between
lying outside and inside the circle, as sketched in
Fig.~(\ref{circles2}).  As it passes through the circle, the residue
of the pole is $\rho(0) = \lim_{\theta\rightarrow 0}\rho(\theta)$.
Thus the integral jumps by $2\pi\rho(0)$.  This is the overlap version
of the Banks-Casher relation;\cite{Banks:1979yr} a non-trivial jump in
the condensate is correlated with a non-vanishing $\rho(0)$.

Note that the exact zero modes related to topology are supressed by
the mass and do not contribute to this jump.  For one flavor, however,
the zero modes do give rise to a non-vanishing but smooth contribution
to the condensate.\cite{Damgaard:1999ij} More on this point later.

\section{Another puzzle}

For two flavors of light quarks one expects spontaneous symmetry
breaking.  This is the explanation for the light mass of the pion,
which is an approximate Goldstone boson.  In the above picture, the
two flavor theory should have a non-vanishing $\rho(0)$.

Now consider the one flavor theory.  In this case there should be no
chiral symmetry.  The famous $U(1)$ anomaly breaks the naive symmetry.
No massless physical particles are expected when the quark mass
vanishes.  Furthermore, simple chiral Lagrangian
arguments\cite{DiVecchia:1980ve,Creutz:2003xu} for multiple flavor
theories indicate that no singularities are expected when just one of
the quarks passes through zero mass.  From the above discussion, this
leads to the conclusion that for the one flavor theory $\rho(0)$ must
vanish.

But now consider the original path integral after the fermions are
integrated out.  Changing the number of flavors $N_f$ manifests itself
in the power of the determinant
\begin{equation}
\int dA\  |D|^{N_f}\  e^{-S_g(A)}.
\end{equation}
Naively this suggests that as you increase the number of flavors, the
density of low eigenvalues should decrease.  But I have just argued
that with two flavors $\rho(0)\ne 0$ but with one flavor $\rho(0)= 0$.
How can it be that increasing the number of flavors actually increases
the density of small eigenvalues?

This is a clear example of how the non-linear nature of the problem
can produce non-intuitive results.  The eigenvalue density depends on
the gauge field distribution, but the gauge field distribution depends
on the eigenvalue density.  It is not just the low eigenvalues that
are relevant to the issue.  Fermionic fields tend to smooth out gauge
fields, and this process involves all scales.  Smoother gauge fields
in turn can give more low eigenvalues.  Thus high eigenvalues
influence the low ones, and this effect evidently can overcome the
naive suppression from more powers of the determinant.

\section{{\AE}thereal instantons}

Through the index theorem, the topological structure of the gauge
field manifests itself in zero modes of the massless Dirac operator.
Let me again insert a small mass and consider the path integral with
the fermions integrated out
\begin{equation}
Z=\int dA\ 
e^{-S_g}\ 
\prod_i (\lambda_i+m). 
\end{equation}
If I take the mass to zero, any configurations which contain a zero
eigenmode will have zero weight in the path integral.  This suggests
that for the massless theory, I can ignore any instanton effects since
those configurations don't contribute to the path integral.

What is wrong with this argument?  The issue is not whether the zero
modes contribute to the path integral, but whether they can contribute
to physical correlation functions.  To see how this goes, add some sources
to the path integral
\begin{equation}
Z(\eta,\overline\eta)=\int dA\ d\psi\ d\overline\psi\  
e^{-S_g+\overline\psi (D+m) \psi +\overline\psi \eta+ \overline\eta\psi}.
\end{equation}
Differentiation (in the Grassmann sense) with respect to $\eta$ and
$\overline \eta$ gives the fermionic correlation functions.
Now integrate out the fermions
\begin{equation}
Z=\int dA\ 
e^{-S_g-\overline\eta(D+m)^{-1}\eta}\ 
\prod_i (\lambda_i+m).
\end{equation}
If I consider a source that overlaps with one of the zero mode
eigenvectors, i.e. 
\begin{equation}
(\psi_0,\eta)\ne 0,
\end{equation}
the source contribution introduces a $1/m$ factor.  This cancels the
$m$ from the determinant, leaving a finite contribution as $m$ goes to
zero.

With multiple flavors, the determinant will have a mass factor from
each.  When several masses are taken to zero together, one will need a
similar factor from the sources for each.  This product of source
terms is the famous ```t Hooft vertex.'' \cite{'tHooft:1976up} While it
is correct that instantons do drop out of $Z$, they survive in
correlation functions.

While these issues are well understood theoretically, they can raise
potential difficulties for numerical simulations.  The usual numerical
procedure generates gauge configurations weighted as in the partition
function.  For a small quark mass, topologically non-trivial
configurations will be suppressed.  But in these configurations, large
correlations can appear due to instanton effects.  This combination of
small weights with large correlations can give rise to large
statistical errors, thus complicating small mass extrapolations.  The
problem will be particularly severe for quantities dominated by
anomaly effects, such as the $\eta^\prime$ mass.  A possible strategy
to alleviate this effect is to generate configurations with a modified
weight, perhaps along the lines of multicanonical
algorithms.\cite{Berg:1992qu}

Note that when only one quark mass goes to zero, the 't Hooft vertex
is a quadratic form in the fermion sources.  This will give a finite
but smooth contribution to the condensate
$\langle\overline\psi\psi\rangle$.  Indeed, this represents a
non-perturbative additive shift to the quark mass.  The size of this
shift generally depends on scale and regulator details.  Even with the
Ginsparg-Wilson condition, the lattice Dirac operator is not unique,
and there is no proof that two different forms have to give the same
continuum limit for vanishing quark mass.  Because of this, the
concept of a single massless quark is not
physical,\cite{Creutz:2004fi} invalidating one popular proposed
solution to the strong CP problem.  This ambiguity has been noted for
heavy quarks in a more perturbative context\cite{Bigi:1994em} and is
often referred to as the ``renormalon'' problem.  The issue is closely
tied to the problems mentioned earlier in defining the topological
susceptibility.

\section{Summary}

In short, thinking about the eigenvalues of the Dirac operator in the
presence of gauge fields can give some insight, for example the
elegant Banks-Casher picture for chiral symmetry breaking.
Nevertheless, care is necessary because the problem is highly
non-linear.  This manifests itself in the non-intuitive example of how
adding flavors enhances rather than suppresses low eigenvalues.

Issues involving zero mode suppression represent one facet of a set of
connected unresolved issues.  Are there non-perturbative ambiguities
in quantities such as the topological susceptibility?  How essential
are rough gauge fields, i.e. gauge fields on which the winding number
is ambiguous?  How do these issues interplay with the quark masses?  I
hope the puzzles presented here will stimulate more thought along
these lines.

\section*{Acknowledgments}
This manuscript has been authored under contract number
DE-AC02-98CH10886 with the U.S.~Department of Energy.  Accordingly,
the U.S. Government retains a non-exclusive, royalty-free license to
publish or reproduce the published form of this contribution, or allow
others to do so, for U.S.~Government purposes.


\end{document}